\documentclass[aps,prl,preprint,superscriptaddress]{revtex4-1}
\usepackage{graphicx}
\begin{document}
%\preprint{}
\title{
Unusual Relationship between Magnetism and Superconductivity in FeTe$_{0.5}$Se$_{0.5}$
}
\author{H.A. Mook}
\email{mookhajr@ornl.gov}
\affiliation{Neutron Sciences Directorate, Oak Ridge National Laboratory, Oak Ridge, Tennessee 37831, USA}
\author{M.D. Lumsden}
\author{A.D. Christianson}
\author{S.E. Nagler}
\affiliation{%
Neutron Scattering Science Division, Oak Ridge National Laboratory, Oak Ridge, Tennessee 37831-6393, USA
}%
\author{Brian C. Sales}
\author{Rongying Jin}
\author{Michael A. McGuire}
\author{Athena Sefat}
\author{D. Mandrus}
\affiliation{
Materials Science and Technology Division, Oak Ridge National Laboratory, Oak Ridge, Tennessee 37831, USA
}
\author{T. Egami}
\affiliation{
Department of Physics and Astronomy, The University of Tennessee, Knoxville, Tennessee 37996-1200, USA
}
\affiliation{
Materials Science and Technology Division, Oak Ridge National Laboratory, Oak Ridge, Tennessee 37831, USA
}
\author{Clarina dela Cruz}
\affiliation{%
Neutron Scattering Science Division, Oak Ridge National Laboratory, Oak Ridge, Tennessee 37831-6393, USA
}%
\affiliation{
Department of Physics and Astronomy, The University of Tennessee, Knoxville, Tennessee 37996-1200, USA
}
\begin{abstract}
We use neutron scattering, to study magnetic excitations in crystals near the ideal superconducting composition of FeTe$_{0.5}$Se$_{0.5}$. Two types of excitations are found, a resonance at (0.5, 0.5, 0) and incommensurate fluctuations on either side of this position. We show that the two sets of magnetic excitations behave differently with doping, with the resonance being fixed in position while the incommensurate excitations move as the doping is changed. These unusual results show that a common behavior of the low energy magnetic excitations is not necessary for pairing in these materials.
\end{abstract}
\maketitle
The discovery of superconductivity above 50K in $R$FeAsF${_x}$O$_{1-x}$ ($R$=rare earth) generated great interest because these materials are the first non-copper oxide superconductors with T${_c}$'s this high \cite{a1,a3}. The discovery immediately brought questions regarding the similarities of these materials to the well known cuprate materials. Typically, superconductivity in both materials arise when a non superconducting parent is doped with holes or electrons. In addition both materials show an excitation in the magnetic spectra known as the resonance \cite{a10,a11,a12}. The resonance excitation scales with the superconducting transition temperature in the same manner for both the Cu and Fe based materials providing evidence that a similar mechanism is responsible for the superconductivity in both cases. Remarkably, the lanthanide oxide fluoride charge reservoir layer found in the original Fe-based superconductors turned out to be unnecessary for superconductivity, and is absent in the FeTe$_{x}$Se$_{1-x}$ systems. However, common to these two systems are the square planer sheets with tetrahedrally coordinated Fe \cite{a4,a8,a5,a6}. Bao et al. \cite{a9} and Li et al. \cite{b9} found antiferromagnetic order in the parent compound $\alpha$-FeTe and that the long-range magnetic order is suppressed by the substitution of Se on the Te sites. Short-range correlations were found to survive in the superconducting phase. A resonance \cite{a10,a11,a12} has been observed in the FeTe$_{x}$Se$_{1-x}$ system   by Mook et al. \cite{a13}, and Qiu et al. \cite{a14}, but the incommensurate excitations were not discovered at that time. Lumsden et. al. later discovered that the wide peaks found by Mook et. al. \cite{a13} resulted from incommensurate excitations and studied them for two concentrations for energies up to about 250 meV \cite{a15}.

The present measurements concentrate on the low energy excitations to study how the resonance and the incommensurate excitations interact. The resonance seems to appear for all the Cu and Fe based materials so it is important to understand its behavior. The stable phase of Fe$_{1+y}$Te$_{x}$Se$_{1-x}$ exists for values of $x$ between 0 and 1.  The Fe(1) sites of the structure are fully occupied, and  excess Fe($y$) partially occupy interstitial sites. For x= 1, the minimum value of y is approximately 0.06, but for smaller values of x (increasing Se content), the value of y approaches 0 \cite{a16,a17}. Large crystals can be grown via the Bridgman method for 0.5$<$x$<$1 \cite{a17}. Although traces of superconductivity with T${_c}$$=$14 K are found in all of the crystals with x$<$0.9, bulk superconductivity is found only in crystals with an average composition near FeTe$_{0.5}$Se$_{0.5}$. Two crystals were used in the experiment. The initial measurements were made on a crystal we denote as A, for which heat capacity, resistivity, and magnetic susceptibility give a T${_c}$$\approx$14 K. Powder x-ray diffraction and EDX analysis of slices cut from various regions of the crystal indicate the crystal consists only of the $\alpha$ phase, but with separate distinct regions. These regions have Te/Se ratios that have more than one composition. Most of the sample has the composition of 55/45 and the rest has less Se. The crystal is thus slightly underdoped; with a composition near 55/45. The exact composition is not needed for this study. Similar composition modulations have been reported in polycrystalline samples \cite{a8}. The second crystal, which we denote as crystal B, has a Te/Se composition that is essentially the ideal value of 50/50 with any composition gradients being small (3$\%$)\cite{a17}.It has a T${_c}$ that bulk measurements show to be slightly higher then that for crystal A, being just under 15K. Heat capacity, resistivity, and magnetic susceptibility measurements are shown in Fig.1a.   Crystal A weighed about 11 grams and crystal B about 15 grams.

\begin{figure}
\includegraphics[scale=.40]{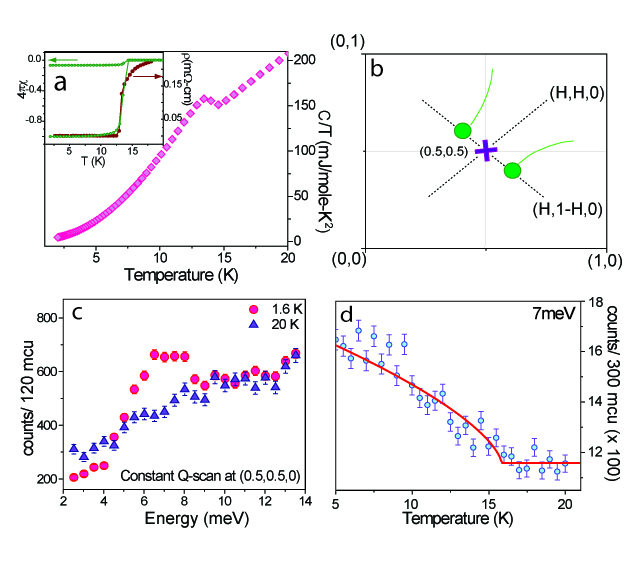}
\caption{
(a) Measurement of the heat capacity of a piece of the FeTe$_{0.5}$Se$_{0.5}$ crystal near and below T${_c}$=14 K, clearly showing bulk superconductivity.  The inset shows the resistivity(right axis) and magnetic susceptibility(left axis) measured with H=20 Oe using zfc and fc protocols. The diamagnetic susceptibility for the zfc data corresponds to complete diamagnetic screening. (b) Diagram of the scattering plane in reciprocal space showing the position of the incommensurate excitations (green circles), the (0.5, 0.5, 0) position as the purple cross, and the scan directions as doted lines. The incommensurate excitations rise quickly in energy. (c) Constant $Q$=(0.5, 0.5, 0) scans taken at 1.6 and 20K. The low temperature curve rises above the high temperature curve the most at about 6 meV giving the resonance energy. (d) Temperature dependence of the scattering at the resonance position showing the rapid increase at T${_c}$. The red solid line is a power law fit to the data.}
\label{figure1}
\end{figure}

The measurements were made at the HB-3 triple axis spectrometer at the High Flux Isotope Reactor, Oak Ridge National Laboratory using collimations of $48^{\prime}$-$40^{\prime}$-$80^{\prime}$-$120^{\prime}$ with a final energy of 14.7 meV using pyrolitic graphite monochromator and analyzer crystals. A pyrolitic graphite filter was placed after the sample to eliminate higher order contamination. A diagram of the scattering plane is shown in Fig. 1b. We use the same tetragonal notation used in the first single crystal measurement made on the Fe based superconductors \cite{a18}, and a diagram is included there showing the difference between the tetragonal and square lattice notation. The magnetically ordered parent compound FeTe was found to order at (0.5, 0, 0.5), which would be positioned at (0.5, 0) on the scattering plane diagram Fig.1b.   As Se is added into the FeTe structure, the scattering at (0.5, 0, 0.5) diminishes and incommensurate excitations are found on either side of the (0.5, 0.5, 0) point, shown by the purple cross. A measurement at 13 meV done using sample A at 1.6K is shown in Fig. 2a. The result shows the incommensurate excitations on either side of the (0.5, 0.5, 0) point. A least squares fit using a gaussian line shape places the excitations at 0.347(7) and 0.657(7) $rlu$ (reciprocal lattice units). Fig. 2b shows the same scan taken with sample B. The incommensurate peaks now further overlap the (0.5, 0.5, 0) position, but can still be fit by gaussian distributions placing the excitations at 0.409(4) and 0.638(4) $rlu$. A fit by a single gaussian gives a quality of fit parameter twice as large. Having the incommensurate excitations come closer to (0.5, 0.5, 0) results in better superconducting properties. The excess Fe also decreases as the Se is increased in crystal B. However, the change is very small and not likely to have a significant effect. A measurement was made on a sample with the composition Fe$_{1.04}$Te$_{0.73}$Se$_{0.27}$ \cite{a15}. The material had steeply rising excitations further out from the (0.5, 0.5, 0) and was non superconducting. This material, having considerably less Se has more excess Fe. However, it is only a 4$\%$ increase compared to ideal 50:50 composition. This further suggests that the position of the incommensurate peaks and superconductivity are related.

\begin{figure}
\includegraphics[scale=.50]{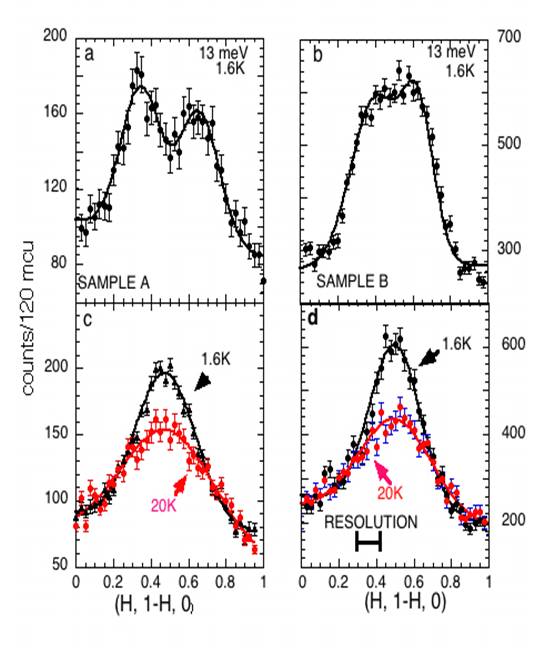}
\caption{
(a) Scan along (H,1-H, 0) at 13 meV for sample A. The scan shows the incommensurate peaks when the measurement was made above the resonance. The line is a gaussian least squares fit. (b) Scan along (H,1-H, 0) at 13 meV for sample B. The incommensurate peaks are much closer to (0.5, 0.5, 0) and overlap considerably. A least squares fit can still be made to two peaks rather then one with a much higher quality of fit parameter. (c) Scan through the resonance in the (H, 1-H, 0) direction at 1.6K and 20K for sample A. The resonance grows out of the 20K peak as the temperature is lowered. However, it only stems from the middle third of the distribution suggesting that it does not come from the incommensurate excitations. (d) Scan through the resonance along (H, 1-H, 0) for sample B. Similar to sample A, the resonance comes out of the middle of the 20K distribution. The resonance is also larger compared to the 20K distribution perhaps reflecting the fact that T${_c}$ is slightly higher. }
\label{figure2}
\end{figure}

Fig.1c shows scans in energy made at (0.5, 0.5, 0) for 1.6K and 20K. This and the measurement in Fig.1d were made with sample B.  There is a large increase in intensity below T${_c}$ at 1.6K at 6 meV giving the resonance \cite{a10,a11,a12}. Typically, the resonance is understood to come from excitations above and below the resonance energy as the temperature is lowered below T${_c}$ \cite{a15,a19}. Here, the intensity above the resonance is unchanged. However, the reduction in the signal within the gap below about 4 meV may account for the increase in the resonance.  In any case, the non-resonance intensity is incommensurate while the resonance appears exactly at (0.5, 0.5, 0). Fig.1d shows that the resonance intensity goes away at T${_c}$.

The incommensurability can be examined by scanning the (0.5, 0.5, 0) position in a direction that goes through the incommensurate peaks. Fig.2c shows a scan taken along the (H, 1-H, 0) direction that, as is shown in Fig.1b, goes through the incommensurate peaks making the pattern much wider.The scan energy used is 6 meV, which is the resonance energy, so that the intensity is much larger below T${_c}$. We see that the extra scattering that goes into the resonance only comes from the center of the 20K distribution. This has not been observed for any other material and suggests that little if any of the increase stems from the incommensurate peaks on either side of the resonance peak. Fig.2d shows a scan at 6 meV along the (H, 1- H, 0) direction for crystal B. This  is observed to be narrower as the incommensurate peaks overlap more. The resonance is seen to be larger compared to the incommensurate peaks. This is expected as the material is closer to the optimum concentration. Density functional calculations have been made for the FeTe$_{x}$Se$_{1-x}$ system and calculations of the electron-phonon coupling show that FeSe is not an electron-phonon superconductor, but within a spin-fluctuation driven picture FeTe with doping would give the highest T${_c}$ of the FeTe$_{x}$Se$_{1-x}$ system \cite{a20}. The calculations give two intersecting cylindrical electron Fermi surfaces at the zone corner which are compensated by lower velocity hole sections at the zone center. Transitions between these Fermi surface sections that have opposite sign order parameters would result from the $s\pm$  pairing state \cite{a20}. These transitions would then appear around the (0.5, 0.5, 0) reciprocal lattice position.  A possible problem with the conjecture that the resonance results from transitions between the two Fermi surface cylinders is that the peak width is only 0.13 $rlu$ wide. The cylindrical sections could differ in shape, size, and three dimensionality.For circular cylinders of radii differing by $\delta$$q$, the scattering will show a high plateau around the nesting vector with width 2$\delta$$q$ \cite{a20,a21}. It would be good to have density functional calculations with the FeTe$_{0.5}$Se$_{0.5}$  composition to more accurately compare the calculated and measured peak widths.

\begin{figure}
\includegraphics[scale=.40]{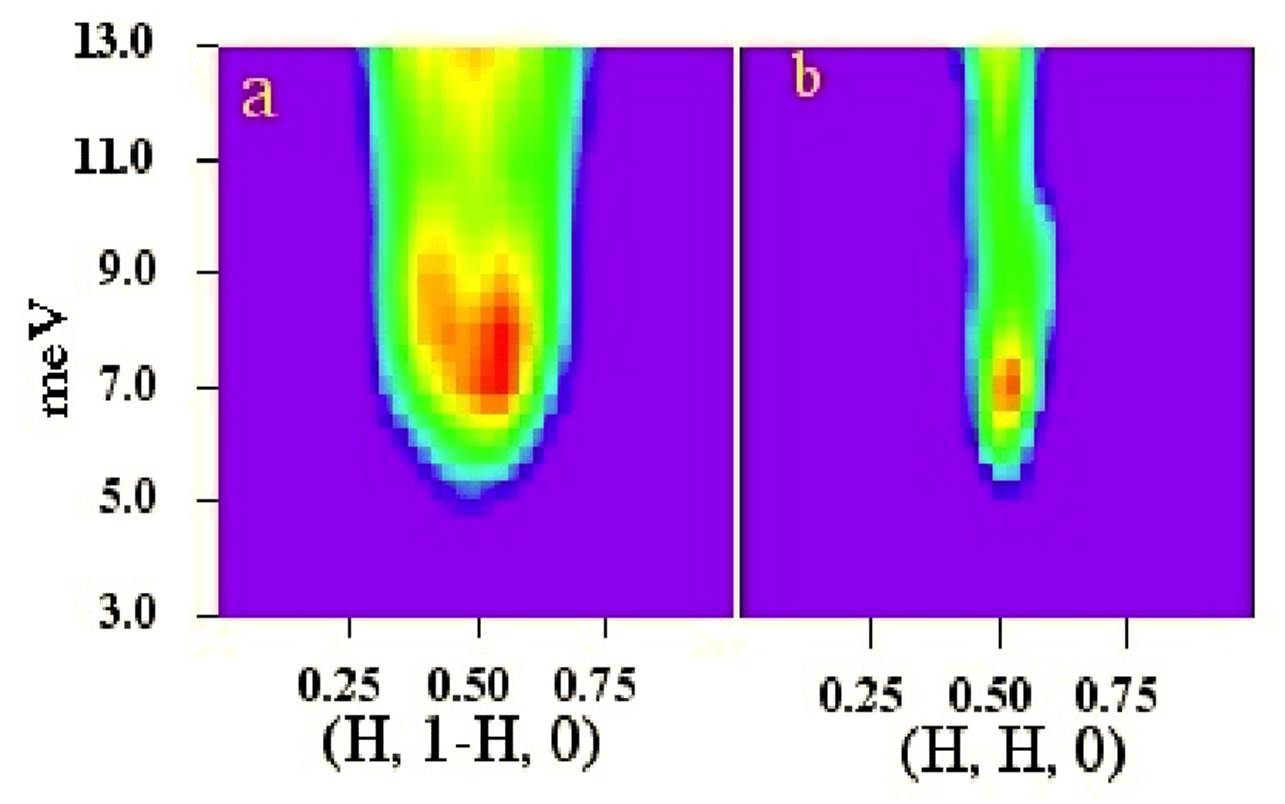}
\caption{
(a)  Contour plot made from scans in the (H, 1-H, 0) direction at 1.6K. It is wide at all energies as the incommensurate peaks are covered as well. The high intensity shown in red comes from the resonance. The map shows that the distribution spreads out slightly as the energy is increased. The lack of intensity at low energies stems from the superconducting gap. (b) Contour plot for (H, H, 0) scans. From 11 meV to higher energies the left hand part of the data could not be obtained due to scattering geometry constraints. The data were thus used from the right side giving a symmetrical pattern. The scattering is narrow at all energies as the incommensurate contribution is largely avoided. }
\label{figure3}
\end{figure}

Fig.3a shows a contour plot of a series of measurements taken along the (H, 1-H, 0) direction through the incommensurate excitations on either side of the (0.5, 0.5, 0). The excitations go nearly straight up in energy and appear to bend slightly outward at higher energies.  The bright spot shows the resonance and the dark area at low energies that results from the gap is easily observable. All the magnetic excitations at this position have a gap below T${_c}$ including the incommensurate ones. Fig. 3b shows a scan along (H, H, 0) which is over the same energy range, but is sharp in $q$. We see the wide variation in width occurs at all energies that we have measured showing the steeply rising nature of the incommensurate excitations.

The measurements show similarities and differences between other magnetically based superconductors. The Cu based superconductors stem from a parent material that shows strong correlations with local moments at the Cu site coupled by superexchange. Doping with either electrons or holes quickly eliminates ordered magnetism in favor of superconductivity. The superconductivity shows $d$-wave symmetry. For the Fe-pnictides, the parent compounds seem to be metallic and show weak magnetism that is sensitive to rare earth substitutions. There is no evidence for gap nodes and thus $d$-wave superconductivity. However, combined hole and electron surfaces results in $s\pm$ order. Charge doping of the parent compound results in superconductivity.

For the FeTe$_{x}$Se$_{1-x}$ materials considered here the behavior is quite different. The doping takes place by means of the substitution of a material that has the same charge. The Fermi surface seems to be the main issue. Adding Se to FeTe changes the magnetic behavior completely and the highest superconductivity occurs with about equal parts Te and Se. The other issue is the observation of steep incommensurate excitations surrounding the (0.5, 0.5, 0) position. Such fluctuations would generally require itinerant electrons and Fermi surface effects. We don't know the relationship between the resonance and the incommensurate excitations. Magnetism and superconductivity seem closely related as superconductivity does not occur until the incommensurate excitations move very near to the (0.5, 0.5, 0) position. The small change in composition between sample A and B does not change the T${_c}$ substantially, but makes a considerable change in the position of the incommensurate peaks. These results show that the interplay between the resonance and incommensurate magnetism with superconductivity is very unusual. We hope that studying the relationship between magnetism and superconductivity in different materials will give us an understanding of the important issues in producing the superconductivity.

We acknowledge helpful discussions with David Singh. Portions of this work were supported by the Scientific User Facilities Division and the Division of Materials Science and Engineering, Office of Basic Energy Sciences, Department of Energy.

\end{document}